\documentclass[aps,prl,reprint,floatfix,amsmath,amssymb]{revtex4-2}
\usepackage{graphicx}
\usepackage{microtype}
\usepackage[hidelinks]{hyperref}
\hypersetup{pdftitle={Milligram-Scale Pendulum Competes with LIGO as a Probe of Collapse Noise at Audio Frequencies},
pdfsubject={Frequency-resolved limits and thermal design of suspended mirrors}}

\newcommand{\prlacklabel}{\textit{Acknowledgments.}---}
\begin{document}
\lefthyphenmin=2 \righthyphenmin=3\relax
\title{Milligram-Scale Pendulum Competes with LIGO\\
as a Probe of Collapse Noise at Audio Frequencies}
\author{Nobuyuki Matsumoto}
\affiliation{Department of Physics, Faculty of Science, Gakushuin University,
1-5-1 Mejiro, Toshima, Tokyo 171-8588, Japan}
\date{November 27, 2026}

\begin{abstract}
Collapse models modify quantum mechanics with a universal noise that destroys superpositions of large objects and shakes all matter with a random force. Because noise with a physical source cannot be exactly white, this force must be tested over a broad range of frequencies. Here we search for this force between 362~Hz and 2~kHz using a 10-s displacement record of a 7.71-mg suspended mirror. For mass-proportional continuous spontaneous localization with a 10-$\mu$m correlation length, our 95\% upper limit on the noise strength is $1.66\times10^{-7}\,\mathrm{s}^{-1}$ for a spectrum flat across 659--718~Hz. In the same band, LIGO strain records give limits more than 20 times higher; its correlated photodiode noise corresponds to about four times our limit. Projections for future milligram mirrors are three orders of magnitude below our present limit.

\end{abstract}
\maketitle

Quantum mechanics allows an object of any size to be in a superposition of two places, and matter-wave interferometers have confirmed this for ever larger objects~\cite{r14}. Whether it remains true for macroscopic objects is not known. Collapse models propose that it does not. In mass-proportional continuous spontaneous localization (CSL), universal noise coupled to mass density suppresses such superpositions faster for heavier objects~\cite{r1,collettpearle}. To an experimenter, this noise acts like an environment that cannot be shielded and whose origin is left open. It has two observable effects: superpositions decay, and every object feels a small random force. Matter-wave interferometers look for the first. A sensitive measurement of the motion of a mirror can look for the second without preparing a spatial superposition~\cite{collettpearle,diosiosc}.

Standard CSL takes the noise to be white, but a physical source cannot produce exactly white noise at all frequencies~\cite{r2,coloredtime}. Searches for spontaneous x-ray emission set the strongest limits on the collapse rate over a broad range of correlation lengths for white CSL~\cite{r15}. Since the spectrum of collapse noise is unknown, extending those limits to audio frequencies requires an assumption about its shape~\cite{r5}.

Existing tests span widely separated frequencies. LISA Pathfinder measured linear and angular acceleration noise at millihertz frequencies~\cite{r3,armanoRot,armano}. A torsion pendulum constrained CSL noise across a range of frequencies, with greatest sensitivity near 5~Hz~\cite{r4}. Measurements near individual mechanical resonances include a levitated sphere at 13~Hz~\cite{r12} and cryogenic cantilevers at 669.7~Hz~\cite{vanE} and above 3~kHz~\cite{r5,r6}. At audio frequencies the gravitational-wave detector LIGO monitors its 40-kg mirrors over a broad band~\cite{r7,r8}.

Nimmrichter, Hornberger, and Hammerer estimated that a milligram-scale suspended mirror could probe lower CSL noise than a 40-kg gravitational-wave detector mirror~\cite{Nimmrichter2014}.
A much smaller mirror can compete with LIGO because collapse noise is correlated only over a short length. For mirrors much larger than this length, axial force-noise power comes mainly from the two faces normal to the motion and scales with their area. Acceleration noise power therefore varies inversely with face area and with thickness squared. For a correlation length of 10~$\mu$m, the same noise gives our 0.5-mm-thick mirror about $2\times10^9$ times as much acceleration noise power as a single 20-cm-thick LIGO mirror.

\begin{figure*}[t]
\centering
\includegraphics[width=\textwidth]{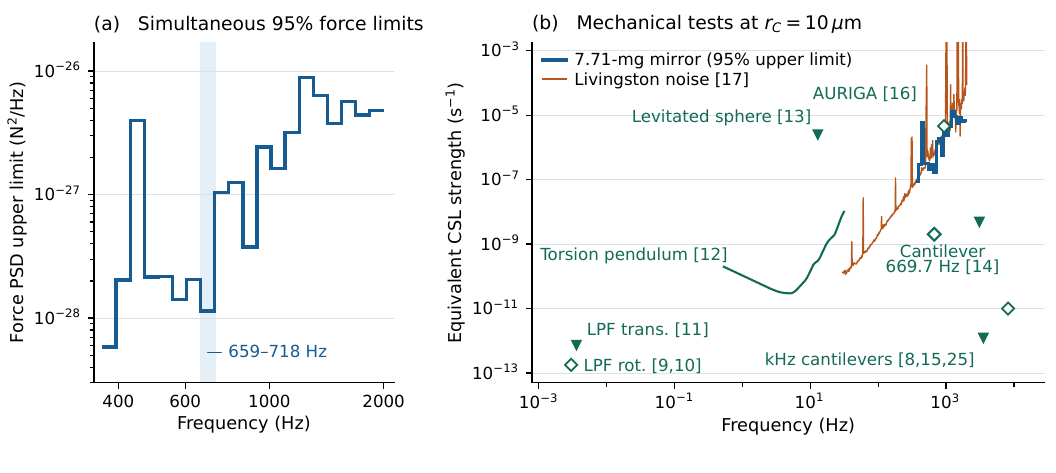}
\caption{(a) Simultaneous 95\% upper limits on force power spectral density in 20 bands. Shading marks 659--718~Hz. (b) Mechanical tests at $r_C=10~\mu$m at their measured frequencies~\cite{r3,armanoRot,armano,r4,r12,vanE,r5,r6,r7,r11}. Blue steps show our flat-spectrum limits. Filled triangles are upper limits, including the LPF translation limit we derive from Ref.~\cite{armano}. Open diamonds and the torsion-pendulum trace show noise equivalents. The 8.17-kHz diamond represents the excess force noise in Ref.~\cite{r11}. LPF denotes LISA Pathfinder. AURIGA is a resonant bar. Orange shows unsqueezed Livingston correlated noise from the fourth observing run (O4)~\cite{r8}.}
\label{fig:limits}
\end{figure*}

Here we analyze a 10-s record from an earlier experiment~\cite{r9,r10,r29} on a 7.71-mg mirror that hangs as a pendulum and is trapped by light. We set upper limits on force noise directly in 20 frequency bands between 362~Hz and 2~kHz, without extrapolation from other frequencies (Fig.~\ref{fig:limits}). We then use LIGO Livingston O4 data to compare the two instruments in matched frequency bands. In the 659--718~Hz band, the 95\% upper limits we derive from LIGO strain records are more than 20 times the upper limit from our pendulum~\cite{r8,gwdata,r18}. The noise correlated between Livingston's two readout photodiodes corresponds to a CSL strength about four times our limit. For a cryogenic cantilever, we infer a noise equivalent about one-eightieth of our flat-spectrum limit, but only within the \mbox{0.044-Hz} linewidth of its \mbox{669.7-Hz} resonance~\cite{vanE}. For Gaussian spectral peaks centered at 580~Hz with widths of 10--50~Hz, our primary limits lie below the corresponding values from all other measurements compared. At 688~Hz, only the \mbox{10-Hz-wide} Gaussian gives this result. \mbox{Finally,} using the suspension loss inferred from ringdown measurements~\cite{r19} and assuming readout at the standard quantum limit (SQL), we project a minimum noise level equivalent to about $1/3000$ of our present upper limit on CSL strength.

\textit{Experiment.}---The mirror hangs as a pendulum and forms one end of an optical cavity~\cite{r9}. Light detuned from the cavity resonance acts as a spring. It raises the resonance of the mirror from 4.53 Hz to about 283.5 Hz, and feedback damps the motion near that resonance~\mbox{\cite{r9,r10}}. The reflected light is recorded as a voltage, sampled at 1 MHz for 10.01 s, and we use the entire record. For an external force $F$, the displacement $x$ of the mirror of mass $M$ obeys
\begin{equation}
 M\big[\ddot x+(\omega_s/Q)\dot x+\omega_s^2x\big]=F,
 \qquad \omega_s=2\pi f_s ,
 \label{eq:response}
\end{equation}
where $f_s$ is the optical-spring frequency and $Q$ the effective quality factor. Because detuning affects both the optical spring and the displacement readout, we convert voltage $v$ to displacement using $x=\alpha_{\rm ref}(283.5~\mathrm{Hz}/f_s)^2v$, with $\alpha_{\rm ref}=2.15\times10^{-10}~\mathrm{m/V}$~\cite{r10}, and infer the force from Eq.~(\ref{eq:response}).

\textit{Upper limits.}---In each of 40 intervals of 250~ms we compute Fourier coefficients of the force. Their summed squared magnitudes give the power in each of 32 logarithmic bands from 130~Hz to 2~kHz. We test for a stationary Gaussian force whose power spectral density (PSD) $S_F$ is flat within the band. The upper limit is the trial $S_F$ for which this force has probability $0.025/32$ of producing a band power at or below the measured value~\cite{r22}. An independent additive background, such as laser frequency noise, can only reduce this probability~\cite{r23}, so the upper limit remains valid without subtracting that background. The same argument excludes any stationary Gaussian force whose PSD exceeds the limit throughout the band. A further probability of 0.025 covers uncertainty in the force calibration, which raises each limit by a factor $1.26^2$. Together these allowances give simultaneous coverage of at least 95\% across the bands, given the response model described next (End Matter).

From the voltage we estimate optical-spring frequencies of 272.4--291.9~Hz. To allow for uncertainty in these estimates, the primary analysis permits any frequency in 265--305~Hz independently in each of 80 or 160 cells that subdivide the intervals (125 or 63~ms). As a check, we widen the optical-spring range to 250--320~Hz. Both ranges allow $250\leq Q\leq2000$ and assume a constant response within each cell. We report the largest upper limit over the allowed parameters and the three time divisions (80 cells, 160 cells and the 25 calibration windows used in Ref.~\cite{r10}). Near the optical-spring resonance this procedure leaves too little force information to set an upper limit, so we obtain limits in 20 bands from 362 Hz to 2 kHz.

\begin{figure*}[t]
\centering
\includegraphics[width=\textwidth]{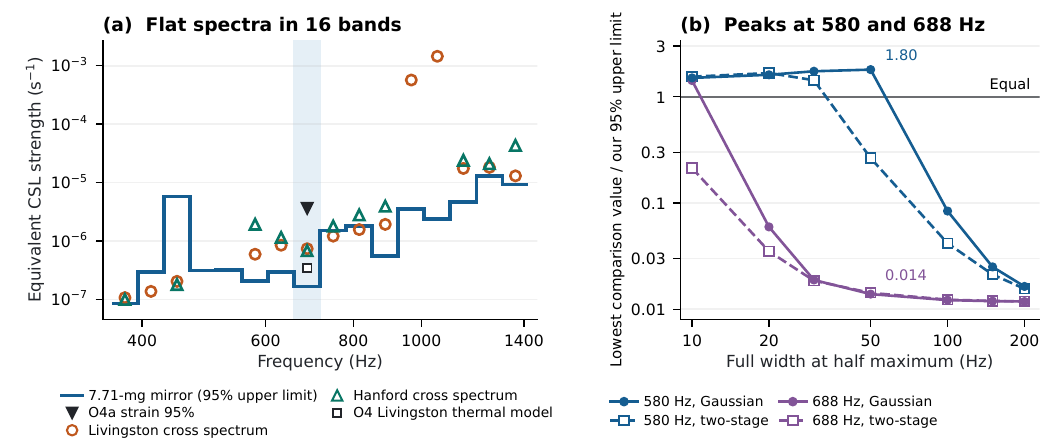}
\caption{(a) The steps show our 95\% upper limits, including calibration uncertainty, in the 16 bands compared with LIGO at $r_C=10~\mu$m. Shading marks the reference band. Circles and triangles show full-band unsqueezed photodiode noise equivalents for Livingston and Hanford, respectively~\cite{r8}. The count of eight bands uses Livingston's quieter half-bands. The square shows Livingston thermal noise modeled in Fig.~13(b) of Ref.~\cite{r8}. The inverted triangle gives the lowest 95\% strain limit from 36 accepted 32-s records in the first part of O4 (O4a), accounting for the records and frequency intervals tested. (b) For each Gaussian and two-stage peak at 580 and 688~Hz, we divide the lowest external upper limit or noise equivalent by our primary upper limit on the peak strength. A ratio above one means that our upper limit lies below every external comparison value. Labels give the 50-Hz Gaussian ratios, set by Livingston's correlated noise at 580~Hz and the 669.7-Hz cantilever at 688~Hz.}
\label{fig:comparison}
\end{figure*}

\textit{From force to collapse noise.}---We assume, as in CSL, a Gaussian spatial correlation of fixed length $r_C$ and a temporal spectrum that does not depend on position. For motion along the mirror normal $z$, the noise then produces the force PSD $S_F(f)=K(r_C)\lambda(f)$. Here $\lambda(f)$ is the CSL strength at frequency $f$, equal to the collapse rate for white CSL, and $K$ depends only on how the mass of the mirror is distributed~\cite{r2,r11}:
\begin{equation}
 K(r_C)=\frac{2\hbar^2r_C^3}{\pi^{3/2}m_0^2}
 \int d^3k\ k_z^2 e^{-r_C^2|\mathbf{k}|^2}
 |\widetilde\rho(\mathbf{k})|^2 ,
 \label{eq:geometry}
\end{equation}
where $m_0$ is the atomic mass unit and $\widetilde\rho(\mathbf{k})=\int\rho(\mathbf{x})e^{-i\mathbf{k}\cdot\mathbf{x}}d^3x$ is the Fourier transform of the mass density $\rho$. For a mirror much wider and thicker than $r_C$, $K$ is proportional to the face area. Our mirror is 3 mm across and 0.5 mm thick. We quote results for $r_C=10~\mu\mathrm m$, where the coating, estimated from the specification, changes $K$ by only about 1\% (End Matter). Dividing a force limit by $K$ gives a limit on $\lambda$. For measured or modeled noise, the \textit{noise equivalent} is the band-averaged force PSD divided by $K$: the flat CSL strength that gives the same band power.

\par
\textit{Limits in 20 bands.}---Figure~\ref{fig:limits} shows the force limits and the corresponding limits on the CSL strength. Our lowest limit is $8.52\times10^{-8}\,\mathrm{s^{-1}}$ in 362--395~Hz, the band most sensitive to the optical-spring model (Table~\ref{tab:response}). For white noise this limit applies at all frequencies. At the same $r_C$, the XENONnT search for collapse-induced x rays constrains white CSL to $3.7\times10^{-13}\,\mathrm{s^{-1}}$~\cite{r15}. Such emission probes the noise near $10^{18}$~Hz, so a spectrum that is not white can evade one test and not the other~\cite{r5}.

For Livingston's correlated noise (described next), we split each band at its logarithmic center and use the quieter half, which favors LIGO. After comparing all 16 bands below 1.42~kHz, we chose 659--718~Hz as the reference band, where this noise equivalent is largest relative to our limit. In this band the force PSD is below $1.14\times10^{-28}\,\mathrm{N^2/Hz}$ and the CSL strength below $1.66\times10^{-7}\,\mathrm{s^{-1}}$.

\textit{Comparison with LIGO.}---LIGO's 40-kg mirrors have lower displacement noise than ours and respond more weakly to the same collapse noise. We convert strain to CSL strength using the four-mirror differential response~\cite{r7} and the mass profiles of the coated mirrors~\cite{r21}.

Measured without squeezing, the cross spectrum of each site's two photodiodes suppresses independent photon-counting noise and retains noise common to both~\cite{r8,r24}. Our 95\% upper limit includes calibration uncertainty and is 2.5 times the mirror's own noise equivalent in the reference band. We compare this limit with the photodiode noise equivalent, which has no assigned confidence level. In an 1800-s Livingston measurement, this noise equivalent is 4.45 times our upper limit over 659--718~Hz [Fig.~\ref{fig:comparison}(a)] and 3.85 times our limit in the quieter half (Table~\ref{tab:ligo}).

From 362~Hz to 1.42~kHz, our upper limit is below Livingston's quieter-half noise equivalent in eight of 16 bands, or in six with the wider optical-spring range. In 555--605~Hz, for example, Livingston's quieter-half noise equivalent is 2.10 times our primary limit. Using the lowest noise equivalent among published photodiode points in each band instead gives six of 16 for Livingston alone and five of 16 when Hanford is included, for either optical-spring range. We omit frequencies above 1.42~kHz from these comparisons because narrow instrumental lines dominate LIGO's band-integrated noise.

With LIGO's nominal strain calibration, 36 public \mbox{32-s} strain records give 95\% upper limits 21.9--36.2 times our reference-band limit~\cite{gwdata,r18}. For a separate Livingston record taken with squeezed light~\cite{r8,gwdata}, \mbox{summing} 1792~s of band power gives a conservative 95\% upper limit 24.4 times ours. The strain readout retains photon-counting noise, which the unsqueezed photodiode cross spectrum suppresses. Integrating separate Livingston strain spectra~\cite{jia} over 659--718~Hz, we find that \mbox{frequency-dependent} squeezing reduced the total noise power by a factor of about three (End Matter). In 430--468~Hz, a single 32-s strain record~\cite{r18} gives a tighter upper limit than ours.

\textit{Comparison with cryogenic resonators.}---Using the reported mode temperature and quality factor of the \mbox{669.7-Hz} cantilever~\cite{vanE}, we estimate a noise equivalent of $2.0\times10^{-9}\,\mathrm{s^{-1}}$, about one-eightieth of our reference-band limit. The resonance linewidth is 0.044~Hz, so this ratio assumes a spectrum flat across the band. For the 713-Hz mode of the NIOBE resonant bar~\cite{tobar}, we use the published strain sensitivity and include only the bar's contribution to the CSL force. The resulting noise equivalent is $4.3\times10^{-7}\,\mathrm{s^{-1}}$, or 2.6 times our reference-band limit (End Matter).

\textit{Peaked spectra.}---We test Gaussian and Lorentzian peaks and the responses of one resonant filter or two in series (single-stage and two-stage), each at seven widths from 10 to 200~Hz. The peaks are centered at 688~Hz in the reference band and at 580~Hz in the 555--605~Hz band. The 580-Hz center lies 90~Hz below the 669.7-Hz cantilever resonance. For each spectrum, an external upper limit or noise equivalent for the peak strength is a \textit{comparison value}. Figure~\ref{fig:comparison}(b) divides the lowest such value by our 95\% upper limit. For the photodiode spectra we take the lowest noise equivalent across integration windows at both sites (End Matter).

At 580~Hz, our primary upper limit is below every comparison value for Gaussian peaks of any width from 10 to 50~Hz and for two-stage peaks with widths of 10, 20, and 30~Hz. For the seven plotted cases with ratios above one, LIGO's correlated noise gives the closest value, 1.5 to 1.8 times our limit, except for the 30-Hz two-stage peak, where the cantilever gives 1.4 times. We repeat the comparison with the wider optical-spring range and with the cutoff for removing voltage offsets and trends lowered from $10^{-8}$ to $10^{-10}$ (End Matter). Our upper limit then remains below all comparison values for five of these seven peaks with the LIGO integration windows and for four with individual points on the published curves. For Gaussian and two-stage peaks at 688~Hz, the cantilever resonance 18~Hz away sets the comparison value for all but the narrowest Gaussian. There, our primary upper limit is below every comparison value only for the 10-Hz-wide Gaussian. The closest is Hanford's correlated-noise equivalent, at 1.42 times our limit (0.823 with the wider optical-spring range and the $10^{-10}$ cutoff). Single-stage and Lorentzian peaks at either center have tails that reach the 3.5-kHz cantilever or LISA Pathfinder, which give lower values.

A limit on the noise in one band does not by itself bound the collapse of a given superposition, which depends on its duration and on the full spectrum~\cite{coloredtime}.

\par
\begin{figure}[t]
\centering
\includegraphics[width=\columnwidth]{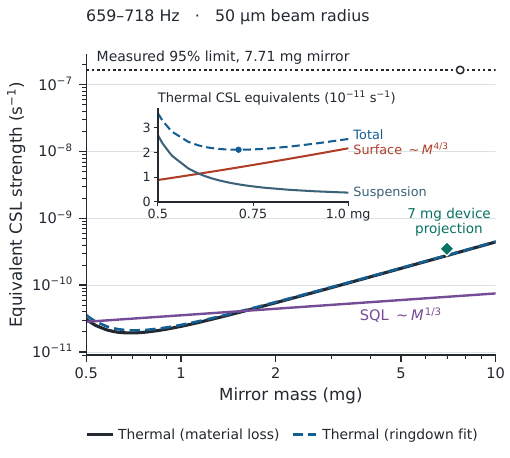}
\caption{Projected CSL noise equivalents in 659--718~Hz for $r_C=10~\mu$m and a Gaussian beam of $50~\mu$m radius ($1/e^2$ intensity). Black and dashed blue show total thermal noise for material loss alone and for the ringdown fit, respectively. Purple shows the SQL. As CSL noise equivalents, mirror-surface noise scales approximately as $M^{4/3}$ and the SQL as $M^{1/3}$, where $M$ is the mirror mass. The inset shows the suspension and mirror-surface contributions to the total thermal noise for the ringdown fit. The dot marks the minimum of their sum. The diamond shows the thermal noise predicted by the same loss model for the existing 7-mg pendulum geometry~\cite{r19} with this beam and coating. The open circle marks the 7.71-mg mirror's measured 95\% upper limit, extended horizontally for comparison.}
\label{fig:design}
\end{figure}

\textit{Prospects.}---Figure~\ref{fig:design} compares two thermal-noise projections for silica disks with diameter three times their thickness, the estimated coating of the present mirror, and a Gaussian beam radius of $50~\mu$m. Both use the fitted stepped-fiber profile of Ref.~\cite{r19} and include mirror-surface Brownian and thermo-optic noise~\cite{r21,designfinite,designto}. The first includes bulk, surface~\cite{designfiber}, and thermoelastic loss in the fiber. The second adds upper-attachment loss fitted to the 7-mg pendulum ringdown.

As the mirror gets lighter, lower fiber tension brings the first violin mode closer to the reference band (2.94~kHz at 7~mg, 0.98~kHz at 0.71~mg). Near the thermal minimum, increasing mirror mass reduces suspension noise and increases mirror-surface noise. Their sum is lowest near 0.7~mg for both loss models. For the loss model fitted to the ringdown, the minimum thermal noise equivalent is $2.1\times10^{-11}\,\mathrm{s}^{-1}$, below the SQL at that mass. With \mbox{readout} at the SQL, the total projected noise equivalent for this loss model reaches its minimum of $5.2\times10^{-11}\,\mathrm{s}^{-1}$ at 0.66~mg. These projections omit technical noise, \mbox{including} laser frequency noise, and assume a coating untested on substrates this thin.

\textit{Conclusion.}---For a spectrum flat across 659--718~Hz and a correlation length of 10~$\mu$m, the 10-s record of our milligram mirror gives a 95\% upper limit of $1.66\times10^{-7}\,\mathrm{s^{-1}}$ on the CSL strength. The noise \mbox{correlated} between Livingston's photodiodes corresponds to a CSL strength 4.45 times this limit, and a separate 1792-s strain record taken with squeezing gives an upper limit about 24 times ours. At the 669.7-Hz resonance of a cryogenic \mbox{cantilever}, we infer a noise equivalent about one-eightieth of our flat-spectrum limit. For~peaks centered at 580~Hz, our primary analysis gives the lowest limits among the measurements compared for Gaussian widths of 10--50~Hz and the tested two-stage widths of 10, 20, and 30~Hz. At 688~Hz only the \mbox{10-Hz} Gaussian gives this result. For readout at the SQL, the ringdown-based projections are about $1/3000$ of our present limit near 0.7~mg and $1/400$ for the existing 7-mg pendulum geometry.

\begin{acknowledgments}
\prlacklabel{}N.M. is supported by JST FOREST Grant No. \mbox{JPMJFR202X}.

This research has made use of data or software obtained from the Gravitational Wave Open Science Center (gwosc.org), a service of the LIGO Scientific Collaboration, the Virgo Collaboration, and KAGRA. This material is based upon work supported by NSF's LIGO Laboratory which is a major facility fully funded by the National Science Foundation, as well as the Science and Technology Facilities Council (STFC) of the United Kingdom, the Max-Planck-Society (MPS), and the State of Niedersachsen/Germany for support of the construction of Advanced LIGO and construction and operation of the GEO600 detector. Additional support for Advanced LIGO was provided by the Australian Research Council. Virgo is funded, through the European Gravitational Observatory (EGO), by the French Centre National de Recherche Scientifique (CNRS), the Italian Istituto Nazionale di Fisica Nucleare (INFN) and the Dutch Nikhef, with contributions by institutions from Belgium, Germany, Greece, Hungary, Ireland, Japan, Monaco, Poland, Portugal, Spain. KAGRA is supported by Ministry of Education, Culture, Sports, Science and Technology (MEXT), Japan Society for the Promotion of Science (JSPS) in Japan; National Research Foundation (NRF) and Ministry of Science and ICT (MSIT) in Korea; Academia Sinica (AS) and National Science and Technology Council (NSTC) in Taiwan.

We used \mbox{GPT-6.2} Astra (OpenAI) and Claude Fable 5.2 (Anthropic) to help write and cross-check the analysis code and to assist with manuscript editing. The author verified the results, reviewed the AI-assisted text, and takes responsibility for the final content.
\end{acknowledgments}

\textit{Data availability.}---The 10.01-s photodetector voltage record used to infer the displacement of the 7.71-mg suspended mirror is available in Ref.~\cite{r29}. The LIGO strain time series are available from the Gravitational Wave Open Science Center~\cite{gwdata,r18,openo4}, and the Livingston strain-noise spectra measured with and without squeezing are available in Ref.~\cite{jia}. For the main LIGO comparison, we digitized the Hanford and Livingston photodiode cross spectra from Fig.~13 of Ref.~\cite{r8}. The accompanying Zenodo archive contains the analysis code, digitized spectra, numerical tables, and reproduction instructions~\cite{thisdata}.

\clearpage
\setlength{\textfloatsep}{7pt plus 2pt minus 2pt}
\setlength{\floatsep}{6pt plus 1pt minus 1pt}
\section*{End Matter}\suppressfloats[t]
\setlength{\baselineskip}{11.8pt}

\looseness=-1\relax \textit{Statistical procedure.}---In each interval we integrate Eq.~(\ref{eq:response}) by parts against Hann-windowed Fourier functions. We form a matrix whose columns are normalized Fourier-coefficient vectors for cell offsets, linear trends and response changes at cell boundaries. From both data and trial signals, we remove components along its left singular vectors when their singular values, relative to the largest, exceed the cutoff of $10^{-8}$. At cutoffs of $10^{-6}$ and $10^{-7}$, all 20 upper limits differ by less than 3\% from their values at $10^{-8}$. At $10^{-10}$, only the 362--395~Hz band loses its upper limit. For the primary optical-spring range, our limit then lies below Livingston's quieter-half noise equivalent in seven of 16 bands instead of eight. Over the full 659--718~Hz band, Livingston's noise equivalent is 4.20 times our limit instead of 4.45. For a frequency band $B$, let $Y$ be the sum of the squared magnitudes of the coefficients remaining after projection. For stationary Gaussian force noise of one-sided \mbox{PSD} $S_F$ constant in $B$ and zero elsewhere, $Y/S_F$ follows $\sum_j w_j Z_j^2$, where $Z_j$ are independent standard normals and $w_j$ are the unit-\mbox{PSD} covariance eigenvalues across frequencies and intervals. The lower quantile $q_p$~\cite{r22} at $p=0.025/32$ gives the limit $Y/q_p$. Within each cell we write the voltage background as $v_0+v_1(t/T-1/2)$, with offset $v_0$ and slope coefficient $v_1$, where $t$ runs from zero to $T$ over each 250-ms analysis interval. For $|v_0|,|v_1|\leq 1~\mathrm V$, the background remaining after projection can raise the upper limits in the two bands of Table~\ref{tab:response} by at most a fraction $3.31\times10^{-7}$ (primary analysis, 160 cells).

\begin{figure}[t]
\centering
\includegraphics[width=0.90\columnwidth]{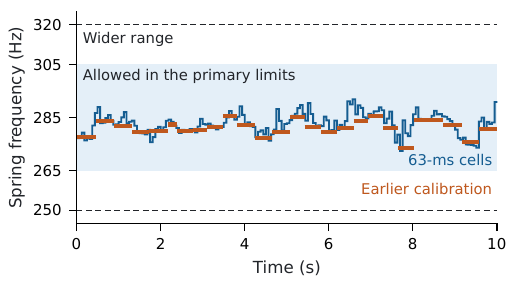}
\caption{Optical-spring frequency tracked from the voltage record~\cite{r29} in 160 cells of 63~ms (thin steps) and in the 25 earlier calibration windows~\cite{r10} (thick segments). Shading and dashed lines show the primary and wider optical-spring ranges, respectively.}
\label{fig:spring}
\end{figure}

\looseness=-1\relax \textit{Optical-spring response.}---Demodulating the reflected voltage near resonance in 63-ms cells gives $f_s=272.4$--$291.9$~Hz (Fig.~\ref{fig:spring}). Our mean frequencies in the 25 earlier calibration windows agree with Ref.~\cite{r10} to 2.2~Hz rms. The 16-Hz resolution leaves the 0.28-Hz linewidth at the adopted $Q=1000$ and its variations unresolved.

\looseness=-1\relax Table~\ref{tab:response} shows the effects of fixing $Q$, widening the frequency range or fixing the response to its tracked values. The tracked row omits tracking uncertainty.

\begin{table}[t]
\caption{Upper limits on the CSL strength at $r_C=10~\mu$m, in units of $10^{-7}~\mathrm{s^{-1}}$, for different treatments of the optical spring. ``Free'' means any value in the range, separately in each cell; the first row gives the primary limits. All rows use the same intervals, projection and calibration.}
\label{tab:response}
\begin{tabular}{llcc}
\hline\hline
$f_s$ in each cell & $Q$ & 659--718~Hz & 362--395~Hz\\
\hline
Free, 265--305~Hz & 250--2000 & 1.663 & 0.852\\
Free, 265--305~Hz & 1000 & 1.662 & 0.849\\
Free, 250--320~Hz & 250--2000 & 2.181 & 1.550\\
Tracked, as known & 1000 & 1.239 & 0.352\\
\hline\hline
\end{tabular}
\end{table}

\looseness=-1\relax The median absolute frequency difference between halves of 63-ms cells is 3.22~Hz, consistent with 999 phase-randomized controls preserving the voltage spectrum (central 95\%, 2.86--3.91~Hz). At $Q=1000$, with the same voltage record and projection, replacing the 63-ms steps in the tracked optical-spring frequency with continuous estimates raised the upper limit by at most 0.32\% in 659--718~Hz and 7.83\% in 362--395~Hz. Four trial modulations of $\pm3$~Hz at one cycle per cell, which the 63-ms frequency estimates do not resolve, changed the reference-band limit by $-1.6\%$ to $+1.8\%$. Faster variations of the actual response remain unconstrained. In the two bands of Table~\ref{tab:response}, no coefficients remain with $f_s$ free in 320 cells. The limits from all three time divisions agree within 9\% in the reference band. In 362--395~Hz, 80 coefficients remain with 63-ms cells, and the limits differ by a factor of two.

\looseness=-1\relax \textit{Coating.}---The front coating has a specified 2-mm diameter and reflectivity above 99.95\% at 1064~nm and $42.26^\circ$ incidence. We model it as 11 quarter-wave $\mathrm{Ta_2O_5/SiO_2}$ pairs, with a single $\mathrm{MgF_2}$ rear antireflection layer. Relative to a uniform disk of equal mass, this stack raises $K$ by about 1\% at $r_C=10~\mu$m and sixfold at 100~nm. The resulting reference-band limit at 100~nm, $2.9\times10^{-4}~\mathrm{s^{-1}}$, depends on the coating model. For nominal coatings and the sampled $r_C$ values (0.1--20~$\mu$m), Livingston's photodiode noise equivalent is 3.9--5.5 times our upper limit over 659--718~Hz.

\looseness=-1\relax \textit{LIGO comparisons.}---Our signal model for the public LIGO strain data includes the 2.2\% loss in reference-band power from downsampling. Of 48 predetermined 32-s records, 36 pass the data-quality and event-separation cuts. Each accepted record's lowest limit accounts for all 48 candidate records, 32 bands and three choices per band (full, lower half, upper half). The tightest limit uses the lower half-band.

\looseness=-1\relax In the reference band, the total strain noise of Jia \textit{et al.}~\cite{jia} gives 72.2 times our limit without squeezing and 23.5 times with frequency-dependent squeezing. This 3.08-fold reduction in total strain power does not determine how squeezing changes the correlation between photodiodes. For squeezed Livingston operation (Fig.~18 of Ref.~\cite{r8}) we sum full-band powers in 56 consecutive 32-s records starting at GPS time 1381903226. The trace of the signal covariance and an upper limit on its largest eigenvalue give a lower bound on the relevant quantile of the summed Gaussian signal power, even for correlated records. The per-test probability $0.05/25\,920$ accounts for bands, full or half bands within individual records, and durations of full-band sums. At 1792~s the noise equivalent is $3.944\times10^{-6}$ and the upper limit $4.058\times10^{-6}\,\mathrm{s^{-1}}$. This limit is 24.4 times our reference-band limit.

\begin{table}[t]
\caption{CSL strengths in 659--718~Hz ($10^{-7}~\mathrm{s^{-1}}$) and ratios to our limit at $r_C=10~\mu$m. L1 and H1 denote Livingston and Hanford. Cross spectra are unsqueezed. The 36 records use strain data~\cite{gwdata,r18}. For modeled thermal noise see Fig.~13(b) of Ref.~\cite{r8}. Limits are at 95\% confidence.}
\label{tab:ligo}
\begin{tabular}{lccl}
\hline\hline
Quantity & Value & Ratio & Kind\\
\hline
Mirror (this work) & 1.66 & 1 & limit\\
O4a strain, tightest of 36 & 36.4 & 21.9 & limit\\
O4a strain, same record & 29.7 & 17.9 & noise\\
L1 strain, squeezed, 1792 s & 40.6 & 24.4 & limit\\
L1 cross, full band & 7.39 & 4.45 & noise\\
L1 cross, quieter half & 6.39 & 3.85 & noise\\
H1 cross, full band & 6.83 & 4.11 & noise\\
H1 cross, quieter half & 6.80 & 4.09 & noise\\
L1 total thermal noise & 3.46 & 2.08 & model\\
\hline\hline
\end{tabular}
\end{table}

\textit{Peak shapes and tests.}---The peaked spectra are $\lambda(f)=\lambda_{\rm pk}h(f)$ for $f>0$, with strength $\lambda_{\rm pk}$ at the peak center $f_c=688$ or 580~Hz. With $g(u)=\exp(-4\ln2\,u^2/W^2)$ and $l(u)=(1+4u^2/W^2)^{-1}$, the Gaussian, Lorentzian, single-stage and two-stage shapes are
\begin{align}
h_{\rm G}&=\frac{g(f-f_c)+g(f+f_c)}{1+g(2f_c)},\\
h_{\rm L}&=\frac{l(f-f_c)+l(f+f_c)}{1+l(2f_c)},\\
h_{1}&=\frac{W^2f^2}{(f_c^2-f^2)^2+W^2f^2},\\
h_{2}&=\left[\frac{a^2f^2}{(f_c^2-f^2)^2+a^2f^2}\right]^2.
\end{align}
\looseness=-1\relax With $a=W/\sqrt{\sqrt{2}-1}$, both $h_1$ and $h_2$ have a full width at half maximum of $W$. For the Gaussian and Lorentzian shapes $W$ is the full width at half maximum of each component. For each peak we use the lowest of five limits. One uses the total power of the band that contains $f_c$. The others coherently combine amplitudes at $f_c$ in groups of 1, 4, 10 or 40 consecutive intervals and sum the group powers. Each test uses $p=0.025/(32\times5)$ for 32 bands and five tests.

\looseness=-1\relax LIGO windows span one, three or five consecutive segments of a published photodiode curve or extend from an interior point halfway to its neighbors in log frequency. No window crosses a gap. We minimize over windows, sites and both linear and log--log interpolation of noise power. Using individual plotted points gives ratios above one for the same spectra at each center in the primary analysis. For 10-Hz Gaussians at 580 and 688~Hz, point ratios are 1.27 and 1.12 (0.843 and 0.650 with the wider optical-spring range and $10^{-10}$ cutoff). In that setting the 10-Hz two-stage ratio at 580~Hz falls from 1.03 to 0.86 using points. For single-stage peaks the lowest comparison value comes from the 3.53-kHz multilayer cantilever~\cite{r6}, or at 688~Hz from the 669.7-Hz cantilever up to 50-Hz width. LISA Pathfinder rotation sets it for Lorentzian peaks at all widths. Its noise equivalent at 3~mHz and the cantilever limit at 3.53~kHz are $1.1\times10^{-6}$ and $7.4\times10^{-6}$ times our reference-band limit.

\looseness=-1\relax \textit{Cryogenic cantilevers.}---For run B~\cite{vanE}, we use the resonance frequency $f_r=669.7$~Hz, mode temperature $T_m=7.7$~mK, and quality factor $Q_r=15\,400$. The one-sided force \mbox{PSD} at resonance is $4k_{\rm B}T_m m_{\rm eff}(2\pi f_r)/Q_r$ with effective mass $m_{\rm eff}=1.5$~ng. For the \mbox{1.51-ng} tip sphere of radius $3.65~\mu$m, Eq.~(\ref{eq:geometry}) gives $K=8.61\times10^{-29}~(\mathrm{N^2/Hz})/\mathrm{s^{-1}}$, omitting the beam and its cross term with the tip. The noise equivalent at resonance is $2.02\times10^{-9}~\mathrm{s^{-1}}$. We evaluate peak shapes at $f_r$ because mode energies were measured within 1~Hz of resonance. For the 50-Hz Gaussian at 580~Hz, the run B value is 63 times our primary limit of $2.41\times10^{-7}\,\mathrm{s^{-1}}$, or 42 times with the Gaussian weighted by the ideal oscillator's power response. The same 580-Hz spectra favor the mirror in both calculations. A revised displacement calibration could imply a run B saturation temperature near 6~mK~\cite{vanE}. At 6~mK the resonance noise equivalent is $1.58\times10^{-9}~\mathrm{s^{-1}}$. At each center the same spectra still favor the mirror in the primary analysis. For the 30-Hz two-stage peak at 580~Hz, the ratio falls from 1.43 to 1.11 (primary) and from 1.05 to 0.819 (wider optical-spring range, $10^{-8}$ cutoff).

\looseness=-1\relax \textit{NIOBE.}---Using the calculated strain sensitivity at 713~Hz in Fig.~12(a) and Eq.~(12) of Ref.~\cite{tobar}, we infer an end-face force-noise amplitude of $6.4~\mathrm{pN}/\sqrt{\mathrm{Hz}}$ over an effective bandwidth of 0.3~Hz. For a uniform niobium cylinder (1510~kg, length $L=2.75$~m, equivalent radius $R=0.143$~m) in the axial mode $\sin(\pi z/L)$, the two end faces give the leading CSL contribution when $r_C\ll R,L$. The noise equivalent at $r_C=10\,\mu\mathrm m$ is $4.3\times10^{-7}\,\mathrm{s^{-1}}$. This bar-only estimate omits radial motion and the readout flap.

\looseness=-1\relax \textit{Assumptions of Fig.~\ref{fig:design}.}---At each mass we solve the prestressed finite-element model with the same fiber profile. For the actual 7-mg pendulum~\cite{r19}, the measured energy-decay rate $2\pi\times361(39)~\mathrm{nHz}$ at 2.63~Hz fixes a loss angle of $2.04\times10^{-3}$ in the upper $200~\mu$m. Holding this angle fixed gives the excess suspension force noise at each mass. At 7~mg, the ringdown fit gives about 4.2 times the suspension force-noise power predicted from material loss alone. We evaluate surface noise separately for each mirror geometry.

\looseness=-1\relax For uncorrelated shot noise and back action, the force PSD at the SQL is $2\hbar/|\chi|$, where $\chi$ is the mechanical susceptibility. The thermal noise is lowest at 0.69~mg for material loss alone and at 0.71~mg for the ringdown fit. For the existing 7-mg pendulum geometry, the ringdown fit gives a thermal noise equivalent of $3.50\times10^{-10}\,\mathrm{s}^{-1}$ or $4.35\times10^{-10}\,\mathrm{s}^{-1}$ including the SQL.

\end{document}